# Long-term Stabilization of Fiber Laser Using Phase-locking Technique with Ultra-low Phase Noise and Phase Drift

Dong Hou, Bo Ning, Shuangyou Zhang, Jiutao Wu, Jianye Zhao

*Abstract*—We review the conventional phase-locking technique in the long-term stabilization of the mode-locked fiber laser and investigate the phase noise limitation of the conventional technique. To break the limitation, we propose an improved phase-locking technique with an optic-microwave phase detector in achieving the ultra-low phase noise and phase drift. The mechanism and the theoretical model of the novel phase-locking technique are also discussed. The long-term stabilization experiments demonstrate that the improved technique can achieve the long-term stabilization for the MLFL with ultra-low phase noise and phase drift. The excellent locking performance of the improved phase-locking technique implies that this technique can be used to stabilize the mode-locked fiber laser with the highly stable H-master or optical clock without stability loss.

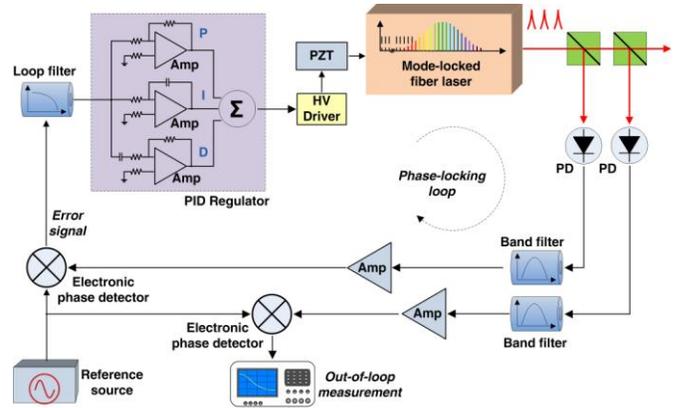

Fig. 1. Configuration of the conventional PZT-tuning-based phase-locking technique with the PID and its measurement. PZT: piezo-electric transducer. HV: high voltage. PD: photo detector. Amp: amplifier.

## I. INTRODUCTION

Ultrafast femtosecond lasers [1] have been widely used in many areas over last years, e. g. frequency metrology, timing distribution, optical communication and optical/microwave frequency synthesis [2]-[5]. In these ultrafast lasers, the mode-locked fiber laser (MLFL) which can generate the shortest optical pulses and ultrastable microwave, has obtained considerable attentions [5], [6]. Among its applications, there is a fundamental demand in achieving both so high and stable repetition frequency of MLFL [7]. To achieve the highly stable MLFL, its repetition frequency should be stabilized to a frequency standard [8], [9]. Many different stabilization approaches have been proposed to achieve this, ranging from the phase-locking schemes where the MLFLs were phase-locked to the highly stable frequency references [10]-[14], to the direct atomic transition locking techniques which were used to direct stabilize the MLFLs to the atomic transitions [15], [16]. Although these approaches have different optical and electronic configurations, their stabilization principle is the stability transfer from the frequency standards to the MLFLs.

Among these stabilization approaches, the phase-locking technique is the most popular and practical stabilization approach, due to its simplicity and economy. A conventional phase-locking technique in the stabilization of the MLFL is shown in Fig. 1. In this phase-locking technique, a standard phase-locked loop (PLL) with an inserted proportional integral derivative (PID) regulator is built between the MLFL and the reference source [17]-[20]. With the conventional phase-locking technique, we can obtain the phase-error signal and feedback it to the piezoelectric transducer (PZT) in the cavity of the MLFL, to phase-lock the repetition frequency of the laser to the electronic reference source in a long time (over several weeks or month).

In the long-term stabilization of the MLFL, the phase noise and stability are the extremely important parameters that evaluate the performance of the locking systems. Therefore, proposing a reliable phase-locking technique with the ultra-low phase noise is a key issue in the design of long-term stabilization system for fiber laser. By stabilizing the MLFL to a very-highly stable frequency standard, e. g. H-master or optical clock with the ultra-low phase noise PLL, we can achieve the ultra-stable fiber laser which will benefit the high-precision frequency metrology, the optical communication and the astronomic combs. However, in the

Manuscript received November 29, 2013; This work was supported in part by the Nature Science Foundation of China (No. 61371074 and 11027404).

The authors are with the Department of Electronics, Peking University, Beijing, 100871, China (e-mail: zhaojianye@pku.edu.cn).

case of ultra-low phase noise stabilization, the current phase-locking technique encounters the difficulties, because the electronic photodetection and phase detection scheme in the conventional phase-locking technique will inevitably suffer from the electronic noises [21], [22]. Therefore, our purpose in this paper is to reveal the phase noise limitation of the conventional phase-locking technique, propose a novel improved phase-locking technique with an optic-microwave phase detector (OMPD), analyze the mechanism of this improved PLL, and conduct a long-term stabilization experiment to prove its ultra-low phase noise performance in the stabilization of the MLFL.

## II. Phase Noise Performances of Electronic phase detection and Optic-microwave phase detection

The conventional phase-locking technique using the electronic phase detection is the most practical approach to stabilize the MLFL. Its configuration is shown in Fig. 1. The principle of the stabilization system is that the repetition frequency of the MLFL is phase-locked to a reference source. In this PLL system, a photodetector is used to convert the optical pulses of the MLFL to electronic harmonic signal. This harmonic signal is first filtered by a band filter, and then mixed with a reference source by an electronic phase detector (EPD) to produces a phase-error signal at the intermediate frequency (IF) port. A loop filter eliminates any unwanted harmonic with higher frequency and generated the Direct Current (DC) phase-error signal. A PID regulator can auto-calculate out the regulation voltage signal to drive the PZT incorporated in the laser's cavity, maintaining the quadrature between the MLFL and the reference source. This auto-tuning technique realizes the conventional PLL between the MLFL and the reference source, which can locks the laser's repetition frequency to that of the reference source with long time. Many works experimentally showed that the conventional phase-locking technique with the EPD is a very effective tool to stabilize the MLFLs. However, the phase-locking technique with the photodetector and the EPD is hard to meet this requirement of ultra-low phase noise and phase drift in stabilizing the MLFL to a highly stable reference source. This is because that in the transfer of timing signal from the optical domain to the electronic domain, the excess phase noise is added in the optical-to-electronic (O/E) conversion process due to nonlinearity, saturation, temperature drift, and amplitude-to-phase conversion in photodiodes, when direct photo-detection is used [22]. In addition, the EPD has the excess phase noise which is introduced by temperature fluctuation, airflow and electronic noise [23]. These excess noises cannot be eliminated completely in the conventional photodetection and phase detection. Here, we will introduce a new optic-microwave phase detection scheme with the ultra-low phase noise to replace the previous photodetection scheme.

The configuration of the optic-microwave phase detection is shown in Fig. 2(a). This phase error detector, which is originally proposed by Kim and Kartner [24], [25], consists of an optical circulator, a 50:50 fiber coupler, a unidirectional phase modulator, and a nonreciprocal quarter-wave bias unit with two Faraday rotators and a quarter-wave plate. A polarization-maintaining (PM) fiber is used to link these optical units to implement a Sagnac loop. The optical pulses from the MLFL are applied to the phase modulator driven by the microwave signal. At the output of the Sagnac loop, the output pulses are modulated with the amplitude proportional to the phase error between the pulse train and the microwave signal. The power difference between the two Sagnac-loop outputs is proportional to the phase error between the optical pulses and the microwave signal. A balanced photodetector is used to detect the power difference, to generate an error voltage signal which can precisely reflect the optic-microwave phase difference. The key issue in implementing the optical phase detector is detecting the phase error between the optical pulses and the microwave signal in the optical domain before the photodetection is involved. In this case, the excess phase noises of photodetector and phase detector are avoided. The function of the phase detection for the OMPD in complex frequency domain is expressed as

$$U_e(s) = GRP_{avg}\phi_o\Theta_e(s) = GRP_{avg}\phi_o(\Theta_i(s) - \Theta_o(s)), \quad (1)$$

where $U_e(s)$ is the phase-error signal. $\Theta_e(s)$ is the phase difference between the two input signals of phase detector. $G$ is transimpedance gain of the balanced photodetector, $R$ is the responsivity of photodiodes, $P_{avg}$ is the average optical power at the two Sagnac-loop output ports, and $\phi_o$ is the modulation depth of the phase modulator. The more details for the OMPD can be found in the reference [24].

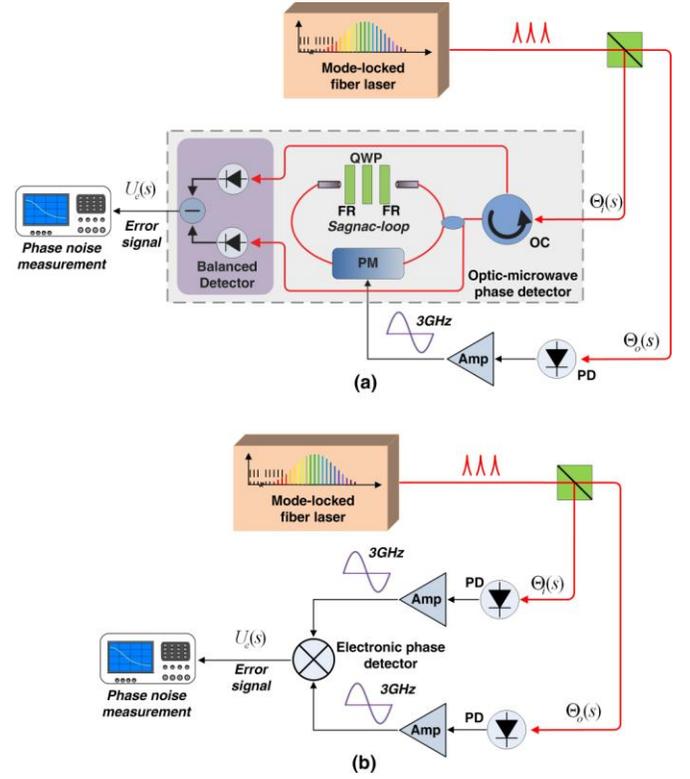

Fig. 2. (a) Phase noise floor measurement with the optic-microwave phase detection. (b) Phase noise floor measurement with the electronic phase detection.

In order to investigate the phase noise performances of the OMPD and EPD, we conducted two phase noise measurement experiments, which can provide the phase noise floors of the two phase detectors. The configurations of the experiments are shown in Fig. 2. A passively erbium-doped MLFL is used as the experimental laser source, and its configuration is similar to that described in the literatures [26], [27]. The MLFL generates an optical beam with the pulses width of approximately 100 fs at a center wavelength of 1550 nm and the 144 MHz repetition frequency. Figure 1(a) shows the optical pulses and the detected coherent microwave (~3 GHz) are delivered in the OMPD, to generate the error signal for the residual phase noise measurement. Figure 1(b) shows the two detected coherent microwaves (~3 GHz) are delivered in a commercial EPD (Hittite, HMC128), to generate the error signal for the residual phase noise measurement. The principle of the measurements is to compare the phases of two coherent signals generated from the same MLFL and achieve the phase noises floors which are mainly introduced by the electronic noise and environmental effects. Figure 3 shows the measurement results of the phase noise floors for the EPD and the OMPD. Curve (i) shows the phase noise floor for the EPD at the 3-GHz carrier frequency. The phase noise floor is -112 dBc/Hz at 1 Hz offset and ~-120 dBc/Hz at 100 kHz offset, respectively, which results in 40 fs rms timing jitter integrated from 1 Hz to 100 kHz. The phase noise floor for the OMPD at the 3-GHz carrier frequency, illustrated by Curve (ii), is -123 dBc/Hz and -140 dBc/Hz at 1 Hz and 100 kHz offset frequency, respectively, which results in 1.4 fs rms timing jitter integrated from 1 Hz to 100 kHz. The phase noise floors results demonstrate that the OMPD will introduce the lower residual phase noise floor, compared to that of the EPD.

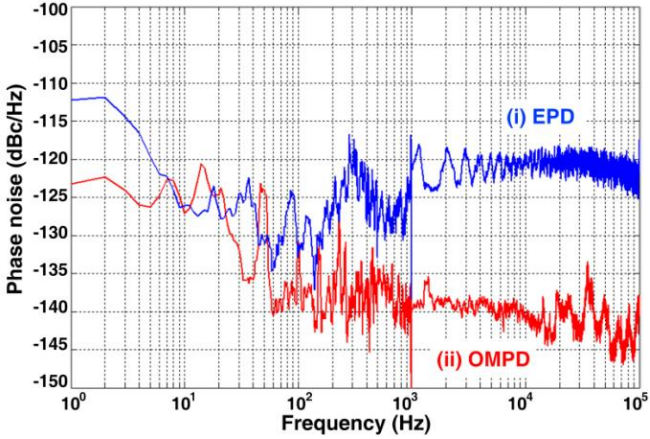

Fig. 3. Phase noise floors for the OMPD and the EPD.

In this section, we analyzed the phase noise limitation of the EPD in the conventional phase-locking technique and demonstrated an optic-microwave phase detection scheme which can break the phase noise limitation and achieve the ultra-low residual phase noise floor. The residual phase noise measurements with the EPD and OMPD proved that the OMPD greatly improved the phase noise performance in the phase detection between the optical pulse and the microwave signal. Therefore, we can replace the EPD by the OMPD in the conventional PLL, to achieve a long-term stabilization system of the MLFL with the ultra-low phase noise. In the next section, based on the optic-microwave phase detection scheme, we will demonstrate a novel improved phase-locking technique in the long-term stabilization of the MLFL with the ultra-low phase noise and phase drift.

### III. IMPROVED PHASE-LOCKING TECHNIQUE USING OMPD WITH ULTRA-LOW PHASE NOISE

In the last section, the phase noise performances of the EPD and OMPD are discussed in detail. The comparison experiments showed that the OMPD provided the ultra-low residual phase noise in the phase detection. Therefore, the OMPD can be used to the potential phase-locking technique in the stabilization of the MLFL with the ultra-low phase noise, by replacing the conventional EPD. However, the previous PZT-tuning-based scheme for controlling the repetition frequency of the MLFL (illustrated in Fig. 1) in the conventional PLL cannot meet the requirement of the low phase noise in the locking loop. This is because that the PZT-tuning-based scheme has two major limitations in the low phase noise stabilization, which are the cavity-length-tuning resolution and the large loop gain of the PLL. These limitations have been discussed in the literature [28]. From the discussion, we find that the PZT-tuning-based scheme will limit the tuning precision of the repetition frequency and the locking-frequency of the PLL. Typically, the locking-frequencies of the stabilization of MLFL using the PZT-tuning-based scheme are limited in low-frequency ($\leq$ 1 GHz), and the minimum phase-tuning resolution is larger than 40 fs. Therefore, if the EPD is replaced by the OMPD, the phase noise of the PLL will be limited by the PZT-tuning-based scheme. In order to meet the requirement of the ultra-low phase noise in the PLL with the OMPD, the conventional PZT-tuning-based scheme for controlling the repetition frequency should be upgraded. In this section, we will first demonstrate a pump-tuning-based scheme in high-accuracy controlling the repetition frequency of the MLFL, for breaking the limitation of the PZT-tuning-based scheme, and then propose a novel improved phase-locking technique using the OMPD with the ultra-low phase noise and phase drift.

*A. Pump-tuning-based scheme in controlling the repetition frequency of the MLFL*

The configuration of the pump-tuning-based scheme [28] for controlling the repetition frequency is shown in Fig. 4(a). This scheme can be used to tune the repetition frequency of the MLFL with more precision by changing the power of its pump source. The MLFL is pumped by a laser diode which is drove by a voltage-controlled current source. With the control of the current source, the pump power is modulated and the repetition frequency of the MLFL is also changed. The voltage-controlled current source, the pump laser diode and the MLFL form a new voltage-controlled oscillator (VCO). Assuming the original cavity length is $L_o$, we have [29]

$$\varpi(t) = \frac{2\pi c}{nL_o} i, \qquad (2)$$

where $\omega(t)$ is the $i^{th}$ repetition radian frequency of the MLFL. $c$ is velocity of light in vacuum. $n$ is the average refractive index

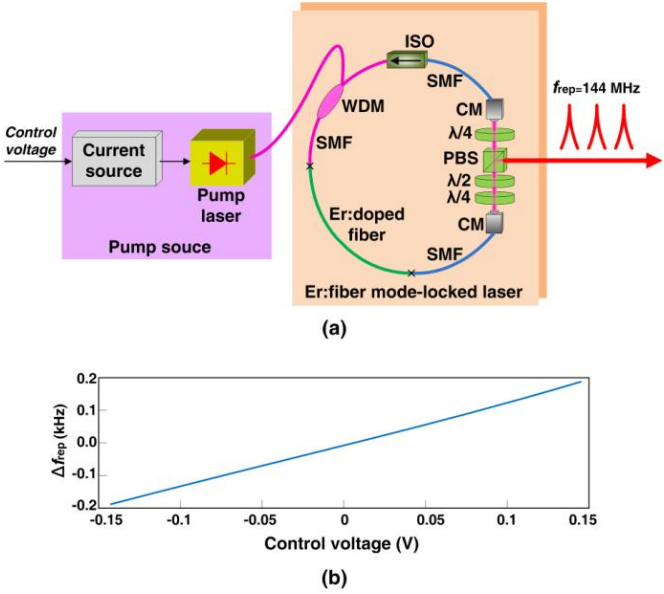

Fig. 4. (a) Pump-tuning-based scheme for controlling the repetition frequency of the Er: MLFL. (b) Experimental repetition frequency curve for the pump-tuning-based scheme. SMF: single-mode fiber. ISO: isolator. CM: collimator. PBS: polarization beam splitter; WDM: wavelength division. multiplexer

of the cavity. This equation shows the nonlinear relationship between $\omega(t)$ and $n$. In order to achieve the transfer function of the pump-tuning-based VCO, a new mathematical model for the VCO should be built. In a MLFL, a certain length of gain fiber is added into the cavity as gain media. The interaction between the atoms of the gain media and electromagnetic radiation affects the MLFL and changes its properties. Therefore, the refraction index of the fiber in the cavity is related to the power of the pump source. The theoretical analysis of the relation between the refraction index and the power of the pump source was discussed in the literature [28], [30], [31] in detail. Here, we rewrite the relationship,

$$\Delta n \propto \frac{3\chi^{(3)}}{4\varepsilon_0 c(1+\chi^{(1)})} \cdot \Delta I, \quad (3)$$

where $\Delta n$ is the change of the refraction index. $\Delta I$ is the changes of the pump's current souce. $\chi^{(1)}$ and $\chi^{(3)}$ are the first-order and third-order susceptibility respectively. $\varepsilon_0$ is the electric permeability of vacuum. The Eq. (3) demonstrates the linear pump current-versus-refraction index relationship. Considering Eq. (2), with a slight change of the refraction index, we have

$$\varpi(t) = \frac{2\pi c}{L_o(n-\Delta n)}i = \frac{2\pi ci}{L_o n}(1-\frac{\Delta n}{n})^{-1} = \frac{2\pi ci}{L_o n}(1+\frac{\Delta n}{n}). \quad (4)$$

Assuming $\Delta n = k_m \Delta I$, $k_m$ is defined as the linear coefficient between the refraction index and current source. Put $\Delta n$ into Eq. (4), we get

$$\varpi(t) = \frac{2\pi ci}{L_o n}(1+\frac{k_m \Delta I}{n}). \quad (5)$$

From the Eq. (5), we have an important conclusion that changing the pump's current source affects the repetition frequency and its high-frequency harmonic in a linear way.

For verifying the linear relationship between the current source and the repetition frequency, we deployed a repetition frequency tuning experiment with the pump-tuning-based scheme. The laser source is an erbium-doped ring-cavity MLFL with the 144.5 MHz repetition frequency, which is same as that described in the last section, whose configuration is also shown in Fig. 4(a). We slightly tuned the control voltage of the pump current (Thorlabs, ITC110) to acquire the changes of the repetition frequency of the MLFL. Figure 4(b) shows that the measured change of the repetition frequency is almost proportional to the tuning change of the control voltage in a linear way. This implies Eq. (5) is correct, and we can use this equation to deduce the VCO model for the pump-tuning-based scheme. The VCO model is given by

$$\Theta_o(s) = \int \omega(t)dt = \frac{2\pi cik_m}{L_o n^2}\int I(t)dt = \\ \frac{2\pi cik_m}{sL_o n^2}I(s) = \frac{2\pi cik_m g_m}{sL_o n^2}U_f(s), \quad (6)$$

where $\Theta_o(s)$ is the phase argument of the MLFL's output signal. $U_f(s)$ is the control voltage of the current source. $g_m$ is defined as the coefficient between the control voltage of the current source and its output current. $2\pi cik_m g_m / L_o n^2$ is the gain of the pump-tuning-based VCO. A literature has experimentally proved that the gain of the pump-tuning-based VCO is much smaller than that of the PZT-tuning-based VCO [28]. Therefore, the pump-tuning-based scheme for controlling the repetition frequency of the MLFL can break the limitation of the PZT-tuning-based scheme described previously. With this pump-tuning-based scheme, we will use the OMPD to design a novel phase-locking system in stabilization of the MLFL with the ultra-low phase noise and phase drift.

### B. The improved phase-locking technique with OMPD

With the help of the pump-tuning-based scheme in accurately tuning the repetition frequency of the MLFL, we can use the OMPD to lock the optical pulses generated from the MLFL to the zero-crossings of the microwave reference source with ultra-low phase noise. The configuration the novel phase-locking technique in the stabilization of the MLFL with the OMPD is shown in Fig. 5. In the PLL with the OMPD, the high-frequency microwave reference signal is used to compare the optical pulses in the OMPD, for generating a phase errors signal. The errors signal is low-pass filtered, amplified, PID-regulated, and fed back to the pump source of the MLFL, for finally achieving the stabilization of the MLFL. The repetition frequency tuning scheme, here, was described in the last subsection. In addition, a drift monitoring circuit is used to detect the slow drift of the input voltage of the pump source. When the pump's power is closed to its limitation, the monitoring circuit can calculate out an appropriate signal to adjust the PZT, compensating the slow drift of the error signal and ensuring the pump source is within the its tuning range. The out-of-loop measurement, shown in Fig. 5, is used to evaluate the locking performance of the pump-tuning-based PLL with the OMPD. The major advantage of this phase-locking technique is that the error signal generated from the OMPD, which avoids the most of electronic noises and environment effects, is a very pure DC voltage signal. With feeding back this pure signal to pump-control the repetition frequency of the MLFL, the stabilization system can achieve the ultra-low phase noise and phase drift.

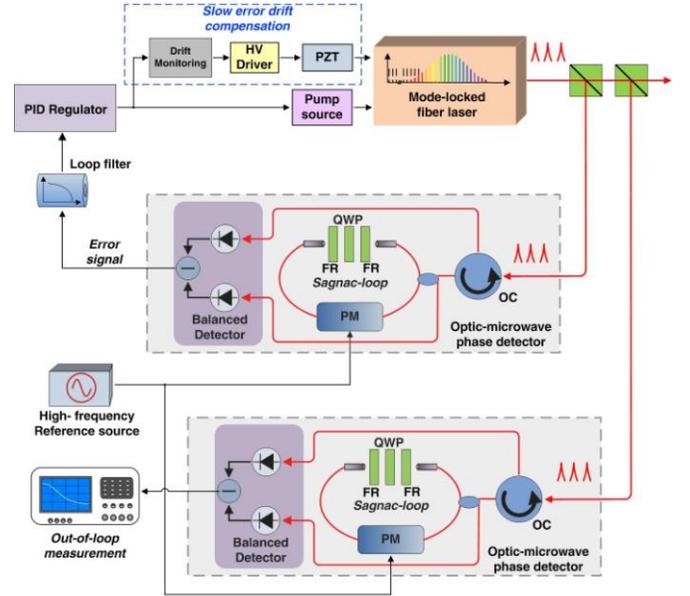

Fig. 5. Configuration of the pump-tuning-based phase-locking technique with the OMPD and its measurement. OC: optical circulator; PM: phase modulator; FR: Faraday rotator; QWP: quarter-wave plate;

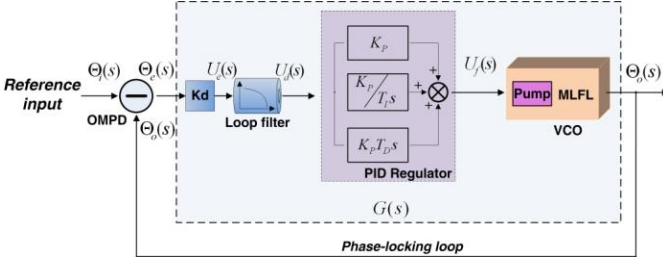

Fig. 6. Block diagram of the pump-tuning-based PLL with the OMPD.

Fig. 6 shows the block diagram of the phase-locking technique with the OMPD. This improved PLL consists of four basic functional elements, which are the OMPD, loop filter, PID regulator and VCO. A PLL is a control system that causes a frequency source to track with another one [32], [33]. More precisely, a PLL is a circuit synchronizing an output signal generated by an oscillator with an input signal in frequency as well as in phase. In order to further understand the mechanism of the improved PLL, a phase-locking model should be built. We are going to first give the functional models of the four elements in the PLL, and then demonstrate the theoretical model of the phase-locking technique with the OMPD. The functional models for the two elements: the OMPD and VCO, which were previously discussed, have been presented in Eq. (1) and (6). Next, we should give the functional models of the other two elements: loop filter and the PID regulator.

*Loop filter.* The phase-error signal generated from the OMPD consists of a number of terms. In these terms, one of the terms is a DC component and roughly proportional to the phase difference $\Theta_e(s)$, which is described in the Eq. (1). The remaining terms are AC components that have the harmonic frequencies. These unwanted higher harmonic must be filtered out via a loop filter. Generally, a low-pass filter is used as the loop filter in PLL. Assuming the transfer function of the loop filter is $F(s)$, the filtered error signal is

$$U_d(s) = F(s)U_e(s), \quad (7)$$

where $U_d(s)$ is the low-pass filtered error signal. $U_e(s)$ is the phase-error signal generated from the OMPD. Equation (7) presents the functional model of the loop filter.

*PID regulator.* A standard PID regulator [34] consists of three elements: proportional, integral and derivative circuits, which are demonstrated in Fig. 1. The proportional circuit can provide an overall control action proportional to the error signal through the all-pass gain factor. The integral circuit can reduce steady-state errors through low-frequency compensation by an integrator, and the derivative circuit can improve transient response through high-frequency compensation by a differentiator. The transfer function of the PID regulator is expressed as [34]

$$\frac{U_f(s)}{U_d(s)} = K_P(1 + \frac{1}{T_I s} + T_D s), \quad (8)$$

where $K_P$ is the proportional gain. $T_I$ is the integral time constant, and $T_D$ is the derivative time constant. $U_d(s)$ is the filtered error signal, and $U_f(s)$ is the control voltage of the current source. The three terms in Eq. (8) are the proportional, integral and derivative terms, respectively. Equation (8) presents the functional model of the PID regulator.

After the functional models of the four elements are given, a linear mathematical PLL model in the stabilization of the MLFL can be developed [33]. We define the open-loop transfer function $G(s)$ as shown in Fig. 6. Based on Eq. (1), (6), (7) and (8), $G(s)$ is given by

$$G(s) = \frac{\Theta_o(s)}{\Theta_e(s)} = k \frac{s^2 T_I T_D + s T_I + 1}{s^2 T_I} F(s). \quad (9)$$

where $k = 2\pi c i k_m g_m G R P_{avg} \varphi_o k_P / L_o n^2$, which is defined as the open-loop gain. Based on the control theory, we define the closed-loop phase-transfer function $H(s)$:

$$H(s) = \frac{\Theta_o(s)}{\Theta_i(s)} = k \frac{s^2 T_I T_D + s T_I + 1}{s^2(k T_I T_D F(s) + T_I) + s k T_I F(s) + k F(s)} F(s). \quad (10)$$

$H(s)$ is relates the input reference source and the laser's output. Generally, the dynamic analysis of a control system is normally performed by means of its closed-loop transfer function $H(s)$. In addition to $H(s)$, an error-transfer function $H_e(s)$ is defined as

$$H_e(s) = \frac{\Theta_e(s)}{\Theta_i(s)} = \frac{s^2 T_I}{s^2(k T_I T_D F(s) + T_I) + s k T_I F(s) + k F(s)}. \quad (11)$$

Equation (9), (10) and (11) present the mathematical model for the improved phase-locking technique with the OMPD. The characteristics of the PLL can be calculated and simulated from these equations. Here, we will omit the calculation and simulation, because the analysis process of this PLL model is almost the same as that of the references [20], [35].

In this section, we describe the defect of the conventional repetition frequency cavity-tuning scheme with the PZT. This defect will limit the application of the OMPD in the ultra-low phase noise phase-locking. To break the limitation of the PZT, we introduced a pump-tuning-based scheme to improve the precision in tuning the repetition frequency. Based on the pump-tuning-based scheme, we proposed a novel phase-locking technique with the OMPD with the ultra-low phase noise. A theoretical model for the proposed PLL was also built. For evaluating the stabilizing performance of the novel phase-locking technique with the OMPD, we deployed a long-term stabilization experiment with this technique. In addition, for comparing the phase noise performances of the PLLs with the EPD and OMPD, a long-term stabilization experiment with the conventional phase-locking technique with the EPD (descripted in Fig. 1) was also conducted. The details of these experiments and measurement results will be discussed in the next Stabilization Experiment section.

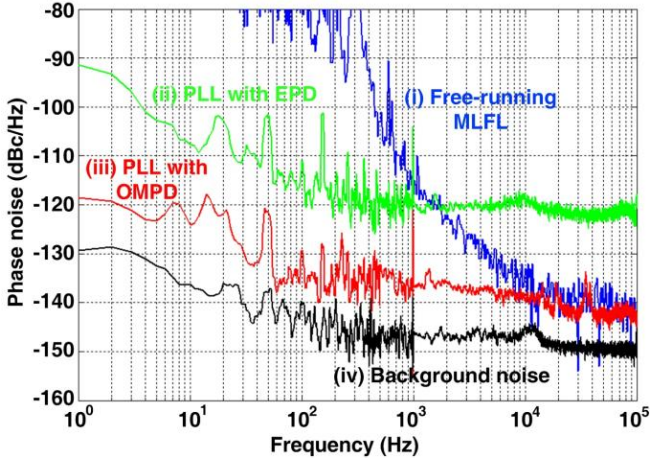

Fig. 7. Phase noises for the free-running MLFL and the stabilized MLFL with the two phase-locking techniques.

## IV. STABILIZATION EXPERIMENT

In last sections, we proposed an improved phase-locking technique with the OMPD in the stabilization of the MLFL. To investigate the performances of the phase-locking techniques with the EPD and OMPD, we conducted two long-term stabilization experiments with these techniques. In our experiments, the MLFL is the same as the previous one, which generated a collimated optical beam with pulses that have duration of approximately 100 fs at a center wavelength of 1550 nm and a repetition frequency of 144 MHz. The goal of the stabilization experiments is to stabilize the MLFL to a highly stable microwave reference source (Agilent, E8257D) using the two phase-locking techniques with the EPD and OMPD, respectively. For the conventional PLL with the EPD, due to the limitation of the PZT's tuning resolution and the VCO's gain, we locked the MLFL's fundamental repetition frequency (144 MHz) signal to the microwave source. For the PLL with the OMPD, we locked its $35^{th}$ harmonic of the repetition frequency (~5 GHz) to the microwave source. All stabilization experiments with the two phase-locking techniques lasted over approximately 10 hours. The out-of-loop measurements, briefly demonstrated in Fig. 1, and Fig. 5, contains the phase noise and phase drift measurements.

### A. Phase noise

Figure 7 shows the phase noise measurement results of the stabilization system with the two phase-locking techniques:

curve (i) shows the absolute single-sideband (SSB) phase noise of the free-running MLFL at carrier frequency of 5 GHz. The phase noise of the free-running MLFL is > -60 dBc/Hz at 1 Hz offset and ~-140 dBc/Hz at 100 kHz offset. With the active phase-locking techniques, the phase and frequency of the MLFL was strictly locked to the reference source. Curve (ii) and (iii) show the out-of-loop residual phase noises of the locked MLFL with the two phase-locking techniques, respectively. The residual phase noise at 144 MHz carrier frequency for the conventional PLL with the EPD, illustrated by Curve (ii), is -93 dBc/Hz and -123 dBc/Hz at 1 Hz and 100 kHz offset frequency, respectively, which results in 45 fs rms timing jitter integrated from 1 Hz to 100 KHz. Curve (iii) shows the residual phase noise at carrier frequency of 5 GHz for the PLL with the OMPD. This phase noise is -118 dBc/Hz and -143 dBc/Hz at 1 Hz and 1 MHz offset frequency, respectively, which results in 2.3 fs rms timing jitter integrated from 1 Hz to 100 kHz. In addition, we also demonstrate the background noise of the OMPD, illustrated by Curve (iv), when only the pulse trains are applied to the OMPD without microwave signal. The experimental phase noise results demonstrate that the improved phase-locking technique with the OMPD can provide the best locking performance in short-term range.

### B. Frequency stability

Figure 8 (a) and (b) show the phase drifts between the microwave reference source and the locked MLFL with the two phase-locking techniques, respectively. The phase drifts data were sampled at 2 samples/s with a 100 Hz low-pass filter.

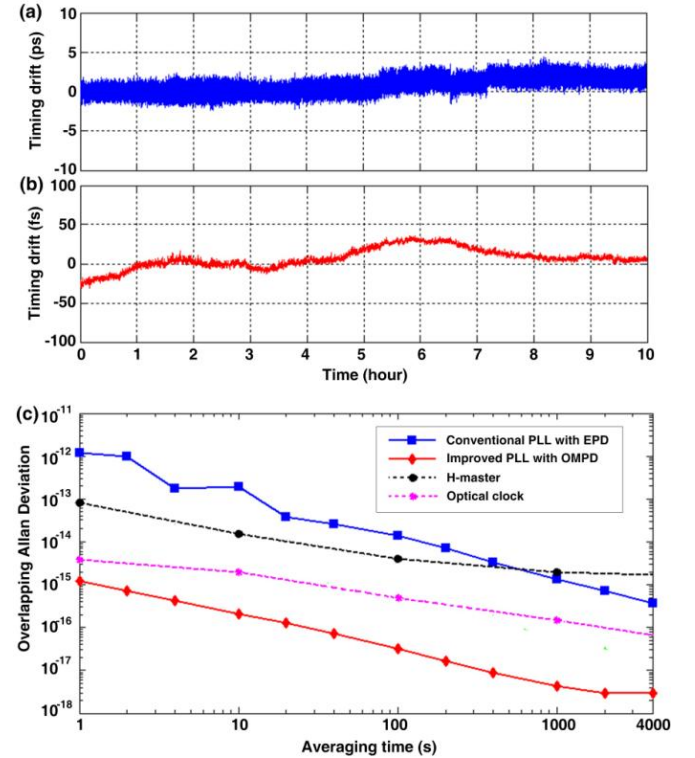

Fig. 8. Phase drifts and frequency stabilities measuremets . (a) Phase drifts for the stabilized MLFL with the two phase-locking techniques. (b) Frequency stabilities for the stabilization system with the two phase-locking techniques, H-master and optical clock.

From these figure it can be seen that the phase drift curves are flat over the measure time (10 hours), which implies that the two phase-locking techniques can both provide the long-term stabilization performances. The integrated rms phase drifts for the conventional PLL with the EPD and the PLL with the OMPD are 1.17 ps and 12 fs, respectively. Based on the measured phase drifts, we also calculated the relative frequency stabilities. Figure 8 (c) shows the relative fractional frequency stabilities of the stabilization system with the two phase-locking techniques. For the conventional PLL with the EPD, the frequency stabilities is $1.2\times10^{-12}$ for a 1-s averaging time and $3.6\times10^{-16}$ for a 4000-s averaging time (filled square in Fig. 8c). For the improved PLL with the OMPD, the frequency stabilities is $1.2\times10^{-15}$ for a 1-s averaging time and $2.9\times10^{-18}$ for a 4000-s averaging time (filled diamond in Fig. 8c). In addition, the stabilities of an H-master clock (filled circle) [36] and an optical clock (star) [37] are also demonstrated in the same figure. Figure 8 demonstrates that the PLL with the OMPD can provide the best long-term stabilization performance. Note that the stability for the conventional PLL with the EPD is larger than that of H-master within 400 s. Therefore, it is difficult to stabilize the MLFL to the highly-stable H-master without the stability loss by using the conventional phase-locking technique with the EPD. In contrast, the relative stability for the PLL with the OMPD is superior to that of H-master and optical clock signal, which implies that we can use the improved phase-locking technique with the OMPD to lock the MLFL to a highly stable H-master or an optical clock signal without stability loss.

## V. CONCLUSION

The long-term stabilization of MLFL using phase-locking technique with ultra-low phase noise is studied in detail. First, we investigated the phase noise limitations of the conventional phase-locking scheme with EPD. To break the limitation of the electronic phase detection, we demonstrated the optic-microwave phase detection in avoiding the electronic noise and environment effects. Next, we improved the conventional phase-locking technique by replacing the EPD with the OMPD, achieving the long-term stabilization of the MLFL with ultra-low phase noise and phase drift. Lastly, we conducted the long-term stabilization experiments with the conventional phase-locking technique with the EPD and the phase-locking technique with the OMPD, respectively, for evaluating their stabilization performance. The experiment results show that the PLL with the OMPD can provide the excellent long-term stabilization for the MLFL. The residual phase noise for the PLL with the OMPD reaches -118 dBc/Hz (-142 dBc/Hz) at 1 Hz (100 KHz) offset frequency, which results in 2.3 fs rms timing jitter, and the long-term (10 hours) phase drift for this phase-locking technique is only 12 fs. This is implied that the novel phase-locking technique with the OMPD can be used to stabilize the MLFL with the ultrastable optical clock signal.

The stabilization scheme with the proposed phase-locking technique with OMPD in this paper can be widely used in some scientific fields, such as timing signal distribution, highly stable optical comb generation and optical-microwave synchronization. The mathematical model we presented in this paper provides a tool to analyze the phase-locking system of the MLFL, and the analysis for the phase-lock technique is also helpful for engineer to design a long-term phase-locking system with ultra-low phase noise in the stabilization of the MLFL.

## VI. ACKNOWLEDGMENT

The authors would like to thank Prof. Zhigang Zhang and Prof. Zhengbin Li, from the department of electronics, Peking University, for helpful discussion and experiment assistant.

**Dong Hou** received the B.Sc. degree in electronic engineering from North China University of Technology, Beijing, China, in 2004, and Ph. D. degree in electronic engineering from Peking University, Beijing, China, in 2012. From 2004 to 2007, he worked in E-world and Lenovo Corporations respectively, as a senior electronic engineer. He is now working in Peking University, Beijing, China, as a postdoctoral fellow.

His current research interests include stabilization technique for mode-locked laser/optical comb with high repetition frequency, highly stable frequency transfer on fiber link, and radio frequency circuit design.

**Bo Ning** received the B.Sc. degree in electronic engineering from Peking University, Beijing, China, in 2008, and Ph. D. degree in electronic engineering from Peking University, Beijing, China, in 2013.

His current research interests include high-stable frequency transfer on fiber link, phase-locking technique for mode-locked laser with high repetition rates and circuit design.

**Shuangyou Zhang** received the B.Sc. degree in college of electronic science & engineering from Jilin University, Changchun, China, in 2010. He is currently pursuing the Ph.D.degree in electronic engineering from Peking University, Beijing.

His current research interests include high-stable frequency transfer on fiber links, stabilization technique for mode-locked lasers, and low noise signal synthesis from optical frequency combs.

**Jiutao Wu** received his B.Sc. degree in electronic engineering from Peking University, Beijing, China, in 2009. He is currently working for the Ph. D. degree in electronic engineering from Peking University, Beijing, China.

He is now engaged in research of ultrafast mode-locked fiber laser and stabilization of the optical frequency comb based on rubidium transitions.

**Jianye Zhao** received his B.Sc. degree in electronic engineering from Fudan University, Shanghai, China, in 1993, and Ph. D. degree in communication engineering from Peking University, Beijing, China, in 1999. He is currently a Professor in the Department of Electronics, School of Electrical Engineering and Computer Science, Peking University, Beijing, China. In 2001, he was pointed as the group leader in the Laboratory of the Circuits and Systems.

Prof. Zhao's research interest is focused on the high-stable timing and frequency distribution in fiber link, long-term phase-locking technique on mode-locked lasers and chaos in circuit network. He is a member of the Chinese Institute of Electronics.